\documentclass[apj]{emulateapj}

\usepackage{bm}
\usepackage{graphicx}
\usepackage{epsf}
\usepackage{graphics}

\shorttitle{Filaments}
\shortauthors{Yan \& Fan}

\begin{document}


\title{Statistical properties of supercluster-like filaments from cosmological simulations}

\author{Heling Yan}
\affil{Department of Astronomy, School of Physics, Peking University, Beijing 100871, China}
\email{yanhl@bac.pku.edu.cn}


\author{Zuhui Fan}
\affil{Department of Astronomy, School of Physics, Peking University, Beijing 100871, China}


\begin{abstract}
In this paper, we study large-scale structures from numerical simulations, paying particular attention
to supercluster-like structures. A grid-density-contour based algorithm is adopted to locate connected groups. 
With the increase of the linking density threshold from the cosmic average density, 
the foam-like cosmic web is subsequently broken into individual supercluster-like groups and further halos.
To be in accordance with normal FOF halos with the linking length of $0.2$ in unit of the average 
separation of particles, halos in this paper are defined as groups with the linking density threshold 
$\rho/\bar \rho=1+\delta=80$, where $\rho$ is the grid density, $\bar \rho$ is the average mass density of the universe.
Groups with lower linking densities are then generally referred to as supercluster-like groups. 
By analyzing sets of cosmological simulations with varying cosmological parameters,  
we find that a universal mass function exists not only for halos but also for 
low-density supercluster-like groups until the linking density threshold decreases to $1+\delta\sim 8$ where the global percolation
of large-scale structures occurs. 
We further show that the mass functions of different groups can be well described by the 
Jenkins form with the parameters being dependent on the linking density threshold. 
On the other hand, these low-density supercluster-like groups cannot be directly associated with the predictions
from the excursion set theory with effective barriers obtained from dynamical collapse models, 
and the peak exclusion effect must be taken into account. 
Including such an effect, 
the mass function of groups with the linking density threshold $1+\delta=16$
is in good agreements with that from the excursion set theory with a nearly flat effective barrier.
A simplified analysis of the ellipsoidal collapse model indicates 
that the barrier for collapses along two axes to form filaments is approximately flat in scales. 
Thus in our analyses, we define groups identified with $1+\delta=16$ as filaments. 
We then further study the halo-filament conditional 
mass function and the filament-halo conditional mass function, and compare them
with the predictions from the two-barrier excursion set theory. 
The shape statistics for filaments are also presented.

%
 
\end{abstract}





\section{Introduction}

One of the key issues in cosmological studies is to understand the physical processes related to 
the structure formation in the universe. In the cold dark matter scenario, gravitational effects
play essential roles in amplifying small density fluctuations generated in the early universe
to shape the large-scale structures seen today. Being directly associated with galaxies and clusters of
galaxies, virialized dark matter halos have been widely studied theoretically and observationally. 
Their mass function, which describes statistically the formation and evolution of dark matter halos,
is shown by numerical simulations to follow a functional form universally valid for a wide range of 
cosmological models \citep[e.g.,][]{she01, jen01}. Such a universality can be largely explained in the context of halo model
which links initial density fluctuations to nonlinear dark matter halos through gravitational collapse models
\citep[e.g.,][]{pre74, coo02}.

Considerable efforts have been made to improve the spherical collapse model to include more realistic 
characteristics in the modeling.  
It has long been realized that the anisotropic features contained in the initial density fluctuations
can be magnified by nonlinear gravity \citep{zel70,zel82}. 
It is expected that the collapse of a region first happens along the direction with the largest eigenvalue of 
the linear deformation tensor, thus leading to a sheet-like structure. Subsequent collapse along the direction of the second 
largest eigenvalue contracts the sheet structure to a filament. A halo can eventually form
once further collapse occurs in the remaining direction. An ellipsoidal collapse model 
is developed to extend these considerations to the nonlinear regime \citep[e.g.,][]{ick73, whi79}. 
The peak-patch scenario further includes the external 
tidal force self-consistently into consideration and improves the modeling of gravitational collapse around initial
density peaks \citep{bon96, bon96b}.
\citet{she01} and \citet{she02} incorporate the peak-patch scenario into the excursion set approach in an averaged way. They first 
obtain statistically the averaged shape parameters of the initial tidal field. These averaged parameters
are then used in the peak-patch ellipsoidal collapse model to derive the collapse criterion.
It is noticed that on average, the halo formation is delayed due to the anisotropy of the gravitational effects. 
The predicted halo mass function (MF hereafter) is then in good agreements with that from numerical simulations.  

\begin{figure*}
\epsscale{1}
\plotone{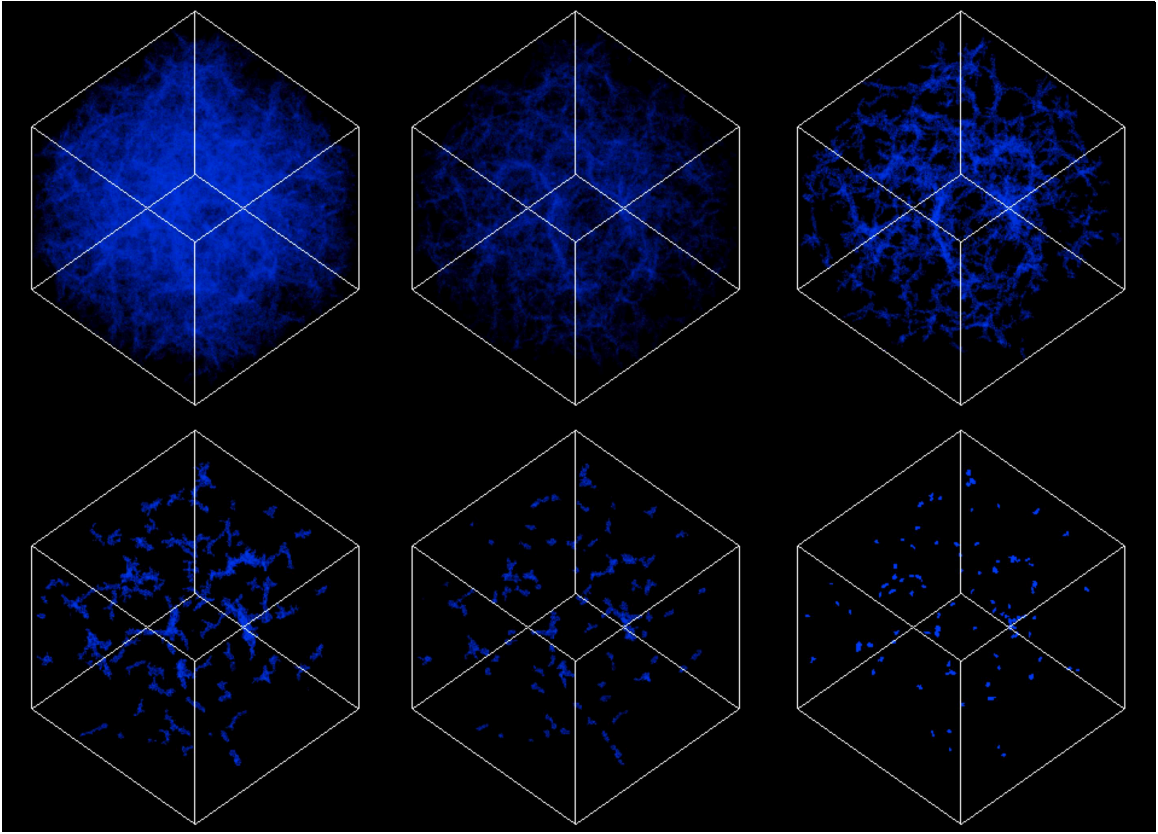}
\caption{Isodensity contours with different linking density thresholds in simulation JS12. 
Only the largest groups (top 100 ranking in mass) are shown. The global cosmic web can already be 
seen in average density contours ($1+\delta=1$) in the top left panel. It gets sharper as 
$1+\delta$ increases to 2 and 4 (top middle and top right panels, respectively). 
At $1+\delta=8$ (bottom left), the cosmic web starts to break out, and large tree structures are seen.
At $1+\delta=16$, the web structure breaks into individual supercluster-like groups (bottom middle). 
Finally at $1+\delta=80$, virialized halos are idenitified (bottom right).}
\end{figure*}

Being very important in the hierarchy of large-scale structures, virialized dark matter halos of galaxy scale and above 
contain only $\sim 40\%$ of the total mass in the universe. Majority of the mass is distributed outside these large halos. 
In the language of halo model, the dominant fraction of the mass in the universe is contained in numerous small halos down to 
very low mass depending on the physical properties of dark matter particles. 
These small halos present anisotropic clustering patterns in space,
and form, together with the massive halos, cosmic web structures. From the view point of the large halos, 
their formation and evolution are affected mainly by the clustering properties of the surrounding small halos as a whole.
Thus to the zeroth order, the mass distribution around a large halo can be described by a smooth component without
considering the individuality of small halos. This approach is clearly stated in the peak-patch scenario \citep{bon96, bon96b}.
In the framework of the excursion set theory, \citet{she06} introduce filaments and sheets to model the large-scale mass distribution 
within which virialized halos are embedded. In their analyses, filaments are treated as an 
intermediate state of the ellipsoidal collapse 
when the collapse finishes along two directions. Then these filaments represent the smoothed version of the 
anisotropic mass distribution around fully collapsed halos.

Various approaches have been proposed to geometrically define filamentary structures 
in cosmological simulations and observations. For example tessellation method is introduced to 
reconstruct the density field, and the edge between tessellations naturally constitutes a segment of filaments 
\citep[e.g.,][]{Ick87,sch00,pla07,dtfe07}. The second order derivatives, namely, Hessian matrix, 
of the tidal field or the density field, is also widely used to classify halos, filaments, sheets and voids according 
to the signs of the eigenvalues of the matrix \citep[e.g.,][]{hah07a,hah07b,sou08,pogosyan09,bon10a,bon10b,ara10}. 
\citet{sto04} propose the so called Candy model for filament finding, 
in which a marked point process with a set of chosen parameters is used to reject 
points at disfavored directions and to locate elongated filamentary segments.
These geometrically defined filaments, however, cannot be directly associated with the excursion-set-based filaments in \citet{she06}.  
The dynamics of long geometrically defined filaments may not be dominated by the local field. Therefore they
may break at the saddle point and accrete into the two ends separately during the late evolution. Furthermore, many geometrical
definitions of filaments concentrate on the features of their spatial distribution rather 
than give rise to countable filamentary objects.  

To emphasize their dynamical structures and to compare with the results of the excursion set theory, in this paper, 
we mainly consider supercluster-like filamentary structures. We adopt a simple but natural definition of filaments by connectivity. 
Specifically, we first obtain the density field on a set of grids from particle positions in a simulation.
Then the site percolation algorithm is applied to link cells together into groups by specifying a 
linking density threshold. At a high enough density threshold, only virialized halos are expected to be identified. 
At lower thresholds, filamentary superclusters surrounding virialized halos are located. The global 
percolation of the cosmic web occurs when the linking density threshold reaches a lower critical value. 
This is illustrated in Figure 1. From top left to bottom right, the linking density thresholds are $\rho/\bar\rho=1+\delta=1, 2, 4, 
8, 16, 80$, respectively. As we will discuss later, the groups identified at $1+\delta=80$ correspond to virialized halos. 
At $1+\delta=16$, the individual structures seen in the plot are related to filaments defined in \citet{she06}.
At $1+\delta=8$, we see the global percolation, and a large structure with a scale comparable to the size of the simulation box
appears. At lower linking thresholds, the global cosmic web gets smoother. It should be noted however, 
that even for $1+\delta=1$, the global web structure can still be seen clearly. 

The paper is organized as follows. \S 2 presents our method in detail. In 
\S 3, we analyze the mass function and the occupation statistics of the identified groups with different linking
density thresholds. In \S 4, we compare our results from simulations with predictions of the excursion set 
theory. Shape statistics are given in \S 5. \S 6 contains summaries and discussions.

\section{Method}

To study the statistical properties of filamentary objects, we analyze sets of 
publicly available numerical simulations from GIF project 
(http://www.mpa-garching.mpg.de.GIF) and Virgo project (http://www.mpa-garching.mpg.de/Virgo), 
which cover a range of cosmological models and simulation parameters.
In addition, we also include in our analyses three $\Lambda$CDM simulations 
kindly provided by Y.P.Jing \citep{jin98}. The relevant simulation parameters are 
listed in Table 1 and Table 2. There are overlaps between the simulations we use
and those analyzed in \citet{jen01} to derive the mass function of dark matter halos, 
and thus comparisons can be made directly between the two studies. 
The simulations we use have relatively low resolutions. 
However, they are sufficient for our purpose of study, which 
aims to investigate large filamentary objects around halos of galactic scale and above 
without concerning the details of individual small subhalos. 



\begin{deluxetable}{ccrrrrrrrrcrl}
\tabletypesize{\scriptsize}
\setlength{\tabcolsep}{0.01in} 
\tablecaption{Simulation Parameters}
\tablewidth{0pt}
\tablehead{
\colhead{Set} & \colhead{Label} & \colhead{Cosmology} & \colhead{$L_{box}$} & 
\colhead{$L_{soften}$} &\colhead{$N_{particle}$}
}
\startdata
JS&JS10&$\Lambda$CDM&100&0.039&$256^{3}$ \\
    &JS11&$\Lambda$CDM&100&0.039&$256^{3}$ \\
    &JS12&$\Lambda$CDM&100&0.039&$256^{3}$ \\
GIF&GIF\_$\Lambda$CDM&$\Lambda$CDM&141.3&0.02&$256^{3}$ \\
    &GIF\_OCDM&OCDM&141.3&0.03&$256^{3}$ \\
    &GIF\_$\tau$CDM&$\tau$CDM& 84.5 & 0.036 & $256^{3}$ \\
    &GIF\_SCDM&SCDM& 84.5 &0.036&$256^{3}$ \\
Virgo&Virgo\_$\Lambda$CDM&$\Lambda$CDM&239.5&0.025&$256^{3}$ \\
    &Virgo\_OCDM&OCDM&239.5&0.03&$256^{3}$ \\
    &Virgo\_$\tau$CDM&$\tau$CDM& 239.5 & 0.036 & $256^{3}$ \\
    &Virgo\_SCDM&SCDM& 239.5 &0.036&$256^{3}$ \\
\enddata
\tablecomments{$L_{box}$ and $L_{soften}$ are in unit of Mpc$h^{-1}$}
\end{deluxetable}

We apply a percolation algorithm to identify connected groups.  
Percolation techniques have been used as group finders ever since the first generation of cosmological 
simulations. The particle based percolation, the FOF algorithm \citep[e.g.,][]{dav85}, which links nearby 
particles by a given linking length, is widely used to locate halos. 
On the other hand, the site percolation \citep[e.g.,][]{bor89,klypin93,shandarin04,shand10}, 
which links adjacent grids 
with density exceeding a threshold, is mainly used to analyze large-scale structures, such as superclusters
and particularly the morphology of global percolative structures at the cosmic average density. 
In this paper, we focus mostly on relatively large structures between virialized halos 
and the global cosmic-average-density surface. We thus choose the latter algorithm, 
which has good enough accuracies for these large structures and can be much faster than 
the particle-based FOF operations. 
The site percolation allows us to find different types of structures by specifying
different linking densities. These structures are thus enveloped by different isodensity 
surfaces, from halos at very high density regions to the global cosmic web at densities approaching 
the average density of the universe (see Figure 1).   

Our specific procedures are as follows. For each simulation snapshot to be analyzed,
we first obtain a density field on a set of regular grids by CIC interpolation from particle positions.
We set a density threshold and pick up only those cells with densities higher than the threshold.
Then a site percolation algorithm, in which cells with shared surfaces are linked together, 
is applied to connect these cells into groups.
For more details of the algorithm, we refer to \citet{new01} \citep[see also][]{klypin93,shandarin04,shand10}.

\begin{figure}
\epsscale{1.0}
\plotone{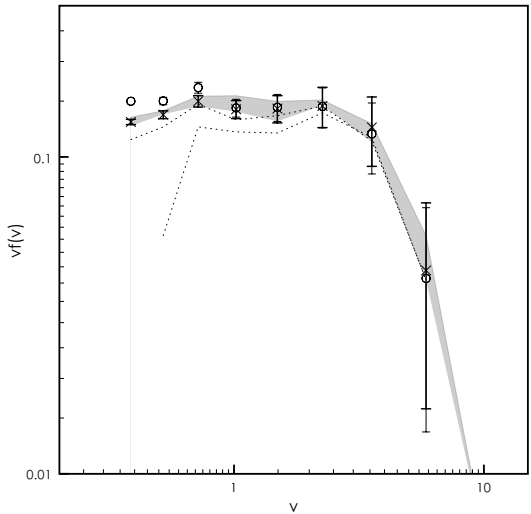}
\caption{Mass functions of site-percolation groups with $1+\delta=80$ (dotted lines) and of FOF halos (circles with 
error bars). For the dotted lines, from top to bottom, the grid number is $512^3$, $256^3$ and $128^3$, respectively.
For the case of $512^3$, the error bars are shown. The shaded region corresponds to the range with $1+\delta$ in between
$72$ and $88$ ($512^3$).}

\end{figure}

\begin{figure}
\epsscale{1.0}
\plotone{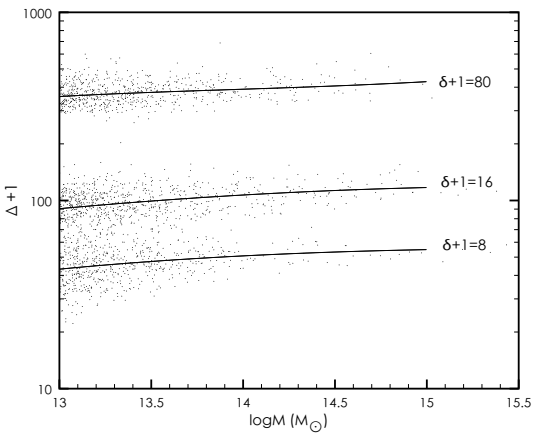}
\caption{Average overdensity of groups for $1+\delta=80$ (top), $1+\delta=16$ (middle) and $1+\delta=8$,
respectively. The horizontal axis is the mass of groups, and the vertical axis is $1+\Delta=<\rho>/\bar \rho$,
where $<\rho>$ is the average density of individual groups, and $\bar \rho$ is the average density of the universe.} 
\end{figure}

If particles distribute uniformly within cells, the site percolation would be equivalent to 
the particle-based FOF with a correspondence between the linking density threshold for grids and the linking 
length for particles. However, particles are not spatially uniform within cells, thus the resolution for the
site percolation depends on the grid size. Compact groups with size smaller than the grid size
can be smoothed and merged artificially into larger groups. To test the grid effects, for 
different grid sizes, we compare halo groups identified with the site percolation to FOF halos with the 
particle linking length of $b=0.2$.
In Figure 2, we show the corresponding scaled mass functions $\nu f(\nu)$ for JS12 simulation, 
where $\nu=\delta_c^2/\sigma_0^2(M)$ and
$ \nu f(\nu)= m^2[n(m,z)/ \bar \rho][d\ln m/ d\ln\nu]$.
Here $n(m,z)$ is the mass function for mass $m$ and redshift $z$. 
The quantity $\sigma_0(M)$ is the rms of the linear density 
fluctuations at the scale corresponding to the halo mass $M$, and is calculated by
\begin{equation}
\sigma_0^2(M)=\frac{1}{(2\pi)^{3}}\int d^{3}\mathbf{k} \tilde{W}^{2}[k,R(M)]P(k),
\end{equation}
where $\tilde{W}[k,R(M)]$ is the Fourier transformation of the top-hat window function with the characteristic scale
$R(M)=[(M/\bar \rho)(3/4\pi)]^{1/3}$, and $P(k)$ is the power spectrum of linear density fluctuations.
The value of $\delta_c$ is taken to be $1.686$. The redshift is $z=0$. 
The symbols are the results for FOF groups. The dotted lines are for the site-percolation groups
with the density threshold $1+\delta=80$, and from top to bottom, the number of grids is $512^3$, 
$256^3$ and $128^3$, respectively. For the upper most dotted line, we also attach the corresponding error bars.
The shaded region represents the range of the mass function with the density threshold
from $1+\delta=72$ to $1+\delta=88$ for the case of $512^3$ grids.
It is seen that at $1+\delta\sim 80$, the mass function of the site-percolation groups agrees
with that of the FOF groups very well at the high mass end. Thus we take $1+\delta= 80$ as the fiducial
threshold for virialized halos. On the other hand, the resolution effect is 
apparent for relatively low mass halos. For the case of $512^3$ grids, halos with $M<10^{12.5} \hbox{ M}_{\odot}$
cannot be well resolved. However, we expect the resolution effect to be weaker for
larger filamentary objects with lower densities, which are our main concerns in the paper. 
Thus we take $512^3$ as our fiducial grid number in the following analyses.


For the site percolation, the density threshold is the critical quantity
to differentiate different groups. It specifies the overdensity level of the envelope of an identified group, and
should be directly associated with the average density within the group. 
Figure 3 shows such a relation, where the vertical axis $1+\Delta$ is, 
in unit of the cosmic density of the universe, the average density  
within individual groups calculated by $M/V$ with $V$ the total volume of a group of mass 
$M$. We present the results for three sets of groups with the density threshold 
$1+\delta=80$, $16$ and $8$ from top to bottom, respectively. The separations of $1+\Delta$ for the 
three sets of groups are clearly seen, with $1+\Delta\approx 400$, $100$ and $50$, respectively.
It is noted that for halos, $1+\Delta\approx 400$, rather than $\sim 200$ defined for spherical halos \citep[e.g.,][]{lac94}. 
This is because the volume of a group calculated here is the actual volume occupied by the connected cells,
and halos are known to be triaxial in shape with a typical value of $\sim 0.5$ for the long-to-short axial ratio
\citep{jin02}. Discussions in \S4 show that groups with $1+\delta=16$ can be related to
filaments defined in the excursion set theory \citep{she06}. We see that they have a typical average overdensity 
of $\sim 100$. For $1+\delta=8$, the global percolation occurs, and such a cosmic web has a typical overdensity
of $\sim 50$. Note that except the grid effect, we do not apply any additional smoothing for the density field in our analyses.
  
\begin{deluxetable}{ccrrrrrrrrcrl}
\tabletypesize{\scriptsize}
\setlength{\tabcolsep}{0.01in} 
\tablecaption{Parameters of Simulation outputs used to fit MF}
\tablewidth{0pt}
\tablehead{
\colhead{Label} & \colhead{z} & \colhead{$\Omega_{0}$} & \colhead{$\Lambda_{0}$} & 
\colhead{$\Gamma$} &\colhead{$\sigma_{8}$} &\colhead{$L_{box}$} &\colhead{$m_{particle}$}
}
\startdata
JS10&0 &0.3 &0.7  &0.21&1.0 &100  &$5.0\times 10^{9}$ \\
JS12&0 &0.3 &0.7  &0.21&1.0 &100  &$5.0\times 10^{9}$ \\
GIF\_$\Lambda$CDM & 0 &0.3   &0.7  &0.21 &0.9  &141.3  &$1.4\times 10^{10}$ \\
						     & 0.5 &0.3 &0.7  &0.21 &0.9  &141.3  &$1.4\times 10^{10}$ \\
						     & 1 &0.3  &0.7  &0.21 &0.9  &141.3  &$1.4\times 10^{10}$ \\
						     & 5 &0.3  &0.7  &0.21 &0.9  &141.3  &$1.4\times 10^{10}$ \\

GIF\_OCDM & 0 &0.3 &0.0   &0.21 &0.85  &141.3  &$1.4\times 10^{10}$ \\
GIF\_SCDM  &0 &1.0 &0.0 &0.5 &0.6  &84.5  &$1.0\times 10^{10}$ \\
GIF\_$\tau$CDM  &0 &1.0 &0.0   &0.21 &0.6  &84.5  &$1.0\times 10^{10}$ \\
Virgo\_ $\Lambda$CDM& 0 &0.3 &0.7  &0.21 &0.9  &239.5  &$6.9\times 10^{10}$ \\
\enddata
\tablecomments{$L_{box}$ is in unit of Mpc$h^{-1}$ and $m_{particle}$ is in unit of M$_{\odot}$h}
\end{deluxetable}

\section{Mass Function of Site-Percolation Groups}

In this section, we analyze statistically the site-percolation groups, 
and present a generalized Jenkins functional form that can describe well the mass function 
of different groups, from virialized halos to low-density supercluster-like groups.   

\subsection{Generalized Mass Function}

It has been shown that to a very high accuracy, the mass function of particle-based FOF 
dark matter halos follows a universal functional form, which is largely independent of 
cosmological models and redshifts \citep[e.g.,][]{she99, jen01}. 
Improved from the original Press-Schechter form \citep{pre74}, two fitting formulae are widely used to 
describe such a universal mass function, namely, the Jenkins form and the Sheth-Tormen form. 

The Jenkins mass function can be written as \citep{jen01}
\begin{equation}
\nu f(\nu)=0.5ae^{-|ln\frac{\sqrt{\nu}}{\delta_{c}}+b|^{c}},
\end{equation}
where $\nu=\delta_c^2/\sigma_0^2(M)$, and $a$, $b$, and $c$ are three parameters with their fitting values 
$a=0.315$, $b=0.61$, and $c=3.8$, respectively, for halos. 

Considering the ellipsoidal collapse model \citep{she01},  \citet{she02} derive the mass function 
from the excursion set theory, which is given by 
\begin{eqnarray}
\lefteqn{\nu f(\nu)=\sqrt{\frac{a\nu}{2\pi}}e^{-a\nu[1+\beta(a\nu)^{-\alpha}]^{2}/2} {}} \nonumber\\
& &{} \times\{1+ \frac{\beta}{(a\nu)^{\alpha}}[1-\alpha+\alpha(\alpha-1)/2+...]\}.
\end{eqnarray}
For dark matter halos, $\alpha\approx 0.615$, $\beta\approx 0.485$, and $a=0.707$. 

Thus as a test for our 
site-percolation-based group analyses, we first study the mass function of dark matter halos
identified with our algorithm. As discussed in \S2, 
our percolation groups with the linking density threshold $1+\delta=80$ correspond well to 
FOF dark matter halos with the linking length $b=0.2$. We then define these groups as 
halos. Figure 4 shows their scaled mass function for all the simulation outputs listed in Table 2.
The solid, dotted and colored lines are the simulation results for $\Lambda$CDM at $z=0$, 
$\Lambda$CDM at $z\neq 0$, and the other cosmological models, respectively. The heavy solid line
is the result of the Jenkins mass function, and the dashed line is for the Sheth-Tormen mass 
function. As expected, the universality of the mass function for the site-percolation halos is 
clearly seen, and both the Jenkins and the Sheth-Tormen functional forms can fit the
simulation results very well. 

\begin{figure}
\epsscale{1.0}
\plotone{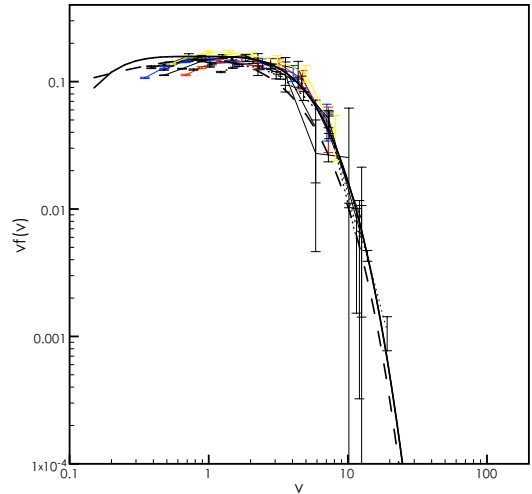}
\caption{Mass functions of site-percolation groups with $1+\delta=80$ for simulations listed in Table 2.
The solid lines are for $\Lambda CDM$ model at $z=0$, and the dotted lines are for $\Lambda CDM$ model at $z\neq 0$
Colored lines show the results of other cosmologies. The heavy solid and dashed lines are Jenkins and 
Sheth-Tormen mass functions for halos, respectively.}
\end{figure}

\begin{figure}
\epsscale{1.0}
\plotone{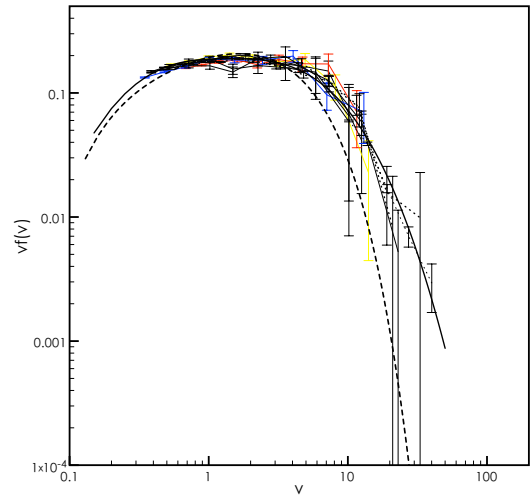}
\caption{The same as Fig. 4, but for $1+\delta=16$. The heavy solid and dashed lines
are fitted mass functions of Jenkins and Sheth-Tormen forms, respectively.}
\end{figure}

We now turn to groups identified with lower linking density thresholds. 
When the density threshold decreases from the halo threshold $1+\delta=80$, 
low density regions surrounding virialized halos are included in groups. 
Nearby halos can also merge into larger filamentary-like objects. When the density threshold approaches
the average matter density of the universe, the global cosmic web extending to the whole simulation box
can be identified. A natural question raised here is whether a universal mass function also exists for 
these supercluster-like groups. To investigate this, for each simulation in Table 2, 
we construct different sets of group catalogs identified with different linking density thresholds,
from the cosmic average density to the density threshold for virialized halos and to even higher thresholds.
For each given threshold, we analyze and compare the mass functions of groups from different simulations. 
It is found that for a wide range of linking density thresholds, the universality of the mass function remains.
In Figure 5, we show the mass functions for groups with the density threshold $1+\delta=16$. The line styles 
for different simulations are the same as those in Figure 4. 
It is seen clearly that the mass function scaled with the quantity $\nu$
obeys a universal form to the level comparable to that of halos. 

To a certain extent, the universality of the mass function for groups beyond halos can be 
understood qualitatively in the same way as for virialized halos. For groups corresponding to 
relatively low linking density thresholds, although they have not been fully virialized, their average densities
are already high enough (see Figure 3) so that their own gravity dominates their dynamical evolution. 
In other words, these groups can be regarded as isolated structures that are in the intermediate stages toward 
forming virialized halos. Considering the process of gravitational collapse of an 
isolated region, as long as all the dependence on cosmology and redshift can be cast into the extrapolated linear
density perturbations, just as in the spherical and ellipsoidal collapse models, 
the process can be described in a universal way. Consequently, a universal mass function for these
groups is expected. On the other hand, it is also expected that the universality of the mass function
should break down when the linking density threshold reaches a low enough level for the occurrence of
global cosmic web structures. We find that the global percolation occurs at $1+\delta\approx 8$, and
indeed the mass function for groups with that linking density threshold and lower
does not show a universal behavior anymore. The global percolation will be discussed in detail in 
\S3.3.   

The universality of mass functions for supercluster-like groups raises a possibility
for us to find an analytical form for mass functions that is generalized from that of dark matter halos.
We consider the Jenkins form of Eq. (2) and the Sheth-Tormen form of Eq. (3).

As seen from Figure 4 and Figure 5, the universal behavior of the mass function depends on the 
linking density threshold. Thus when we fit the functional forms to the simulation results,
we expect that the best fitted values for the parameters in Eq. (2) and Eq. (3) are functions
of the linking density threshold.  

We first consider the Sheth-Tormen form [Eq. (3)]. It involves three parameters $a$, $\alpha$, and $\beta$.
From the excursion set theory \citep[e.g.,][]{she02,she06}, the parameters $\alpha$ and $\beta$
are related to the shape of the collapse barrier with respect to $\nu$, and $a$ reflects the 
overall height of the barrier. For dark matter halos, $a$ is usually taken to be $a=0.707$ 
\citep[e.g.,][]{she01}. In \citet{she06}, they extend the ellipsoidal collapse model to obtain 
the respective collapse barriers for filamentary and sheet-like objects. Their derived barriers for different
types of objects are different only in parameters $\alpha$ and $\beta$. To be in accordance
with their analyses, in our fitting here, we fix $a=0.707$ and vary $\alpha$ and $\beta$. In the next section,
we will consider more general fitting to further discuss the relation between our results and the 
excursion set theory. 

The heavy dashed line in Figure 5 shows our best fit result with Eq. (3) for
groups with $1+\delta=16$. It is seen that the two-parameter ($\alpha$ and $\beta$)
Sheth-Tormen functional form  can fit the low-mass end of the mass function rather well.
At high mass end, however, the model gives a poor fit to the simulation results. This
indicates that the simple excursion set theory cannot apply directly to low-density groups. 
As we will discuss in the next section, this should be related to the well known peak-exclusion 
effect \citep{bon96}. For low-density groups, such effect is stronger than that for
high-density halos, and thus the deviation between the Sheth-Tormen fitting and the simulations
is more apparently seen in Figure 5 than that in Figure 4 for halos. 

For Jenkins form of Eq. (2), we regard it as an empirical form, and thus
all the three parameters $a$, $b$ and $c$ are treated as free parameters in our fitting. 
We then find that Eq. (2) can fit the mass function of groups with different
linking density thresholds very well. We further obtain a generalized fitting for the
three parameters that is applicable to all the groups in consideration, from halos
with the linking density $1+\delta=80$ to low-density groups with $1+\delta>8$.  
This is given by
\begin{equation}
a=0.5852(1+\delta)^{-0.1562}
\end{equation}
\begin{equation}
b=0.1898(1+\delta)^{0.2701}
\end{equation}
\begin{equation}
c=1.927(1+\delta)^{0.1529}.
\end{equation}
For halos with $1+\delta=80$, we have $a=0.295$, $b=0.620$ and $c=3.766$, 
in agreement with the original fitting of \citet{jen01} $a=0.315$, $b=0.61$ and $c=3.8$.
The slightly lower value of $a$ is due to the grid effect in our site-percolation analyses. 
For $1+\delta=16$, we have $a=0.380$, $b=0.401$ and $c=2.944$, and the corresponding
fitting is shown by the heavy solid line in Figure 5.

The generalized Jenkins mass function obtained here allows us to perform statistically 
the abundance analyses not only for halos but also for more extended low-density groups.
Its cosmological applications will be explored in our future studies. 

\subsection{Occupation Statistics}

Several weak lensing measurements reveal the existence of massive dark clumps with 
unusually high mass-to-light ratios \citep[e.g,][]{erb00,ume00,mah07}. 
One proposed explanation is that those dark clumps may arise from the projection effect of low-density filaments
with their elongations happening to be near the line of sight. 
Galaxies within these low density areas are thought to be less clustered than that in high density 
clusters of galaxies of similar mass. However, the proper question concerned in the dark clump problem
should be whether a filament has significantly less projected number of galaxies than that of a cluster of the same projected mass. 
In other words, it is more or less the total number of 
galaxies contained in a filament or in a cluster that matters.   

Numerical studies show that the gravitational effects determine dominantly the occupation statistics of galaxies in a halo
\citep[e.g.,][]{kra04}. Thus the subhalo occupation statistics can give important information on the occupation distribution of galaxies.
Here we compare filament occupation distribution (FOD) of subhalos with  halo occupation distribution (HOD) of subhalos. 
For FOD, we define groups picked up with the linking density threshold $1+\delta=16$ as filaments, 
in accordance with the definition of \citet{she06} (see \S4). For each filament, we count the number of
halos inside it. As discussed previously, halos with $M<10^{12.5}\hbox{ M}_{\odot}$ cannot be
well resolved with our site-percolation group finder due to the grid effects.
Thus here for occupation analyses, we use the particle FOF method with the linking 
length parameter $b=0.2$ to find halos in filaments.
For high-density virialized halos, we do not directly count the subhalos inside them because the 
simulations we used have limited dynamical resolutions. 
Instead, we adopt the HOD fitting result 
for the average number of subhalos with minimum mass $M_{min}$ from \citet{kra04}, which is given by 
\begin{equation}
\langle N\rangle=1+(\frac{M}{M_{1}}-C)^{\beta_{s}},
\end{equation}
where $M$ is the host halo mass, $M_{1}/M_{min}=22$, $C=0.045$ and $\beta_{s}=1.03$. 

\begin{figure}
\epsscale{1.0}
\plotone{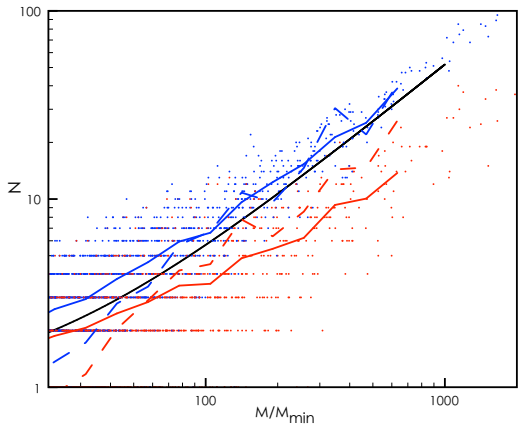}
\caption{Occupation statistics for filaments with $1+\delta=16$. The results from simulations JS10, JS11, and
JS12 are shown. Red dots show the number of halos in individual filaments. Blue dots are for the
results with subhalos added in. The red and blue solid lines are the average results of the red and blue dots,
respectively. The black solid line is the subhalo HOD result in virialized halos. 
The red and blue dashed lines are the second moments of the red and blue dots, respectively.}
\end{figure}

The results of the occupation distribution are presented in Figure 6. The horizontal axis is for the 
mass $M$ of the host filament/halo in unit of $M_{min}$. We limit our analyses 
to $M\ge 10^{12}\hbox{ M}_{\odot}$ in accordance with the relatively low numerical resolutions of the simulations used.
For the same reason and also concerning galactic-scale subhalos, we take $M_{min}=10^{12}\hbox{ M}_{\odot}$.
The FOD results for individual filaments are shown by red dots. The red solid line shows the average value 
of $N$ from the red dots.  
The average HOD from Eq. (7) for virialized host halos is
shown as the black solid line. Naive comparison between the black and red solid lines indeed
leads to the conclusion that the number of subhalos in a high-density virialized halo is statistically larger than 
that contained in a low-density filament of the same mass. It should be noted, however,
that we find halos in a low-density filament by particle FOF group finder with $b=0.2$. 
These halos can have mass well above that of the typical galactic halo, and thus
are expected to further contain subhalos of galactic scale in them. Those subhalos
cannot be adequately identified in our simulation analyses due to the limited dynamical resolutions.
On the other hand, in the studies of \citet{kra04},
they use high resolution simulations
and their halo finder can resolve well subhalos and even sub-subhalos. 
Thus the red solid line and the black solid line in Figure 6 cannot be directly comparable. 

To make a more meaningful comparison between
the occupation statistics of virialized halos and that of filaments,
we need to add subhalos into the halos in filaments. Then for each halo found in a filament,
we adopt Eq. (7) as the average value to randomly assign a number of subhalos to it. 
The corresponding modified FOD results are shown in blue dots in Figure 6. The blue solid 
line is the average of the blue dots.
We see that the blue solid line lies above the black solid line, showing that
after taking into account subhalos, the average FOD result is actually larger by $\Delta N\sim +1$
than that of HOD of the same mass. Such a difference may be understood as follows. Considering
two large halos in a low-density filament with one of them being the largest halo in the filament.
When the filament evolves further to form a virialized halo, the less massive halo
is very likely to merge into the largest one and loses its identity, thus reducing the number of
occupation by $\Delta N=1$. Therefore if there is a proportional relation between the subhalo FOD/HOD and galaxy
FOD/HOD, the mass-to-light ratio for a filament is comparable and can be even lower than 
that of a cluster of the same mass, leading to difficulties for the filament interpretation of dark clumps.
On the other hand, although close relations between the occupation distribution of
subhalos and that of galaxies are expected, differences between the two can exist. Thus
detailed analyses of galaxy occupation distribution are further needed concerning the quantitative interpretation of 
dark clumps with filaments. 
  
In Figure 6, the red and blue dashed lines are the second moments, defined as
$\sqrt{<N(N-1)>}$, of the distributions of the red and blue dots, respectively. 
The red dashed line lies above the red solid line, showing the super-Poisson behavior
for the FOD in the case without adding subhalos into halos in filaments. Considering subhalos in halos, 
the blue dashed line is nearly the same as the blue solid line, and the distribution of the 
blue dots is consistent with the Poisson distribution. 
This is because we add in subhalos assuming the Poisson distribution in accordance with
the HOD analyses of \citet{kra04}.

\subsection{Critical Phenomenon}


The large-scale connected cosmic web is the most striking feature seen in numerical simulations 
as well as in large galaxy redshift surveys. Both the power spectrum of initial density 
perturbations and late-time nonlinear gravitational interactions play important roles in 
shaping the cosmic-web structure \citep[e.g.,][]{bon96b, shand10}. From Figure 1, 
we can see that with the linking density threshold being the cosmic average density, 
i.e., $1+\delta=1$, the global web structure is clearly seen. As the linking density threshold
increases, the cosmic web becomes sharper. At $1+\delta\sim 8$, the global web structure 
starts to break out into large tree structures with massive halos in their central regions. 
With further increased linking density thresholds, the large tree structures break into 
individual groups dominated mainly by their local gravity. 
Thus at relatively low linking density thresholds, 
the percolation groups have two subclasses, relatively isolated ones in low-density 
regions, and the large connected groups that account for over $\sim 50\%$ of volume occupied by
all the groups \citep[e.g.,][]{shand10}. While the local gravity should still 
dominate the formation of isolated groups, 
nonlocal effects affect the formation of the global web structure significantly. 
Therefore there should exist a critical linking density threshold below which the
universality of the mass function of groups breaks down due to the nonlocal gravitational effects
on those large tree-shaped groups.
 

\begin{figure}
\epsscale{1.0}
\plotone{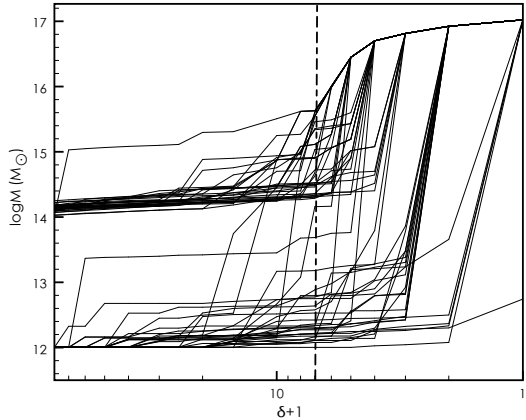}
\caption{Merging path along $1+\delta$ for halos of $M\sim10^{14} M_{\odot}$ and
$M\sim10^{12} M_{\odot}$, respectively.}
\end{figure}

To understand the global percolation phenomenon, in Figure 7, we show the merging path with respect to the linking 
density threshold for halos of $M\sim 10^{14}M_{\odot}$ and $M\sim 10^{12}M_{\odot}$, respectively.
The results are for JS12 simulation. It is seen that, for massive halos, they more or less stay
as isolated ones until $1+\delta\sim 8$. After that, these large groups 
merge into the largest structure, i.e., the cosmic web. This merging process
is rather sharp. At $1+\delta\sim 4$, all the massive halos merge into the cosmic web,
and no isolated ones are left. On the other hand, for relatively low mass halos, 
their merging path in $1+\delta$ space is extended. At $1+\delta>8$, they gradually merge into
larger individual groups with the decrease of the linking density thresholds. At $1+\delta\sim 8-4$, 
some of them merge into the cosmic web. However, there are still isolated halos left even at $1+\delta=1$. 
The difference in the merging path between the massive and low mass halos reflects the fact 
that all massive halos locate at high density regions which eventually become parts of the global web. 
For low mass halos, while some of them are in high density regions, a considerable fraction of them
are in low density void regions and can keep as individual groups at $1+\delta \sim 1$. 

Figure 8 shows the scaled mass function for all the simulations listed in Table 2. The linking density
threshold is $1+\delta=8$. As expected, we can see that the mass function for 
different simulations behaves differently at high mass end, in contrary to those shown in Figure 5
with $1+\delta=16$. The differences are larger than the expected
Poisson fluctuations, showing that the universality of the mass function does not hold
anymore due to the formation of large tree structures which later connect to form the global cosmic web. 
Thus approximately, $1+\delta=8$ is a critical value for the linking density threshold 
below which the global web structure starts to be apparent, resulting the breakdown of the
universality of the mass function for groups.  
 


\begin{figure}
\epsscale{1.0}
\plotone{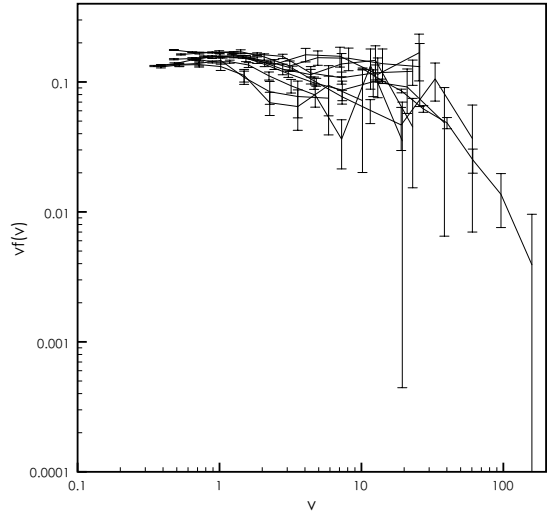}
\caption{Mass functions for groups with $1+\delta=8$. The results for simulations listed in Table 2 are shown.}
\end{figure}

\section{Relation With the Excursion Set Theory} 
\subsection{Unconditional Mass Function}

Originated from \citet{pre74}, the halo formation theory has linked virialized halos 
to linear density fluctuations through dynamical collapse models. 
The halo mass function can then be statistically determined by specifying
a proper collapse barrier for linear density fluctuations smoothed over suitable scales. 
To overcome the cloud-in-cloud problem, 
\citet{bon91} propose the excursion set theory, which considers trajectories in $R$-domain
of the linear density perturbations at random spatial positions. Here $R$ is the smoothing scale applied to 
smooth the linear density perturbation field. By relating the mass fraction contained in halos
with mass greater than $M$ to the volume fraction occupied by trajectories 
first crossing the specified barrier at scales larger than $R(M)$, 
the excursion set theory can give rise to the halo mass function with 
the correct "fudge factor" of $2$ \citep{bon91}. In this theory, all the nonlinear gravitational effects
are encoded in the shape of the barrier, which in turn is determined by dynamical collapse models.
The simple spherical collapse model leads to a constant barrier that is independent of the halo mass $M$ or 
equivalently the smoothing scale $R$.
Taking into account the ellipsoidal collapse, \citet{she01} obtain a $M$-dependent moving barrier 
for the halo formation, which leads to a better agreement between the derived mass function and that 
from N-body simulations. The moving barrier for trajectories of linear density perturbations can be written as
\begin{equation}
B[\sigma_0^2(M), z]=\sqrt{a}\delta_{c}(z)[1+\beta(a\nu)^{-\alpha}],
\end{equation}
where $z$ is the redshift, $\delta_c$ is the barrier in the spherical collapse model, 
and $\alpha\approx 0.615$ and $\beta\approx 0.485$. 
The corresponding mass function from the excursion set theory is then approximately determined by Eq. (3)
\citep{she02}. It is noted that from the ellipsoidal model, we should have $a=1$. However, studies 
show that $a\approx 0.7$ is often required to be in agreement with the mass function from 
numerical simulations \citep[e.g.,][]{she01}. 

\citet{she06} further extend the ellipsoidal collapse model to consider
the formation of sheet-like, filament-like and halo structures, defined as
having collapsed along one, two and all three axes, respectively. 
The effective barriers for the three classes of objects can be generally described by Eq. (8)
with  
\begin{equation}
\alpha\approx 0.55, \beta\approx -0.56 \hbox{ }\hbox{ for sheet},
\end{equation}
\begin{equation}
\alpha\approx 0.28, \beta\approx -0.012 \hbox{ }\hbox{ for filament},
\end{equation}
and 
\begin{equation}
\alpha\approx 0.61, \beta\approx 0.45 \hbox{ }\hbox{ for halo}.
\end{equation}
It is seen that for halos, the effective barrier increases monotonically 
with $\sigma_0(M)$, delaying the formation of low-mass halos due to the nonspherical 
collapse. For sheet-like objects, it is a decreasing function of $\sigma_0(M)$.
It is for filaments that the barrier is nearly a constant in $\sigma_0(M)$.  

It has been extensively shown that the excursion set theory with the moving barrier
and the adjusted $a$ parameter can give rise to the mass function of virialized halos
that fits the simulation results very well \citep[e.g.,][]{she01, she02} (see also Figure 4).  
To a certain extent, it is expected that the excursion set theory with
suitable barriers should also be able to 
model the mass function of low-density groups as long as their formation 
is dominated by their local gravity. The sheets and filaments discussed in \citet{she06}
are among these low-density groups.  
The universality of the mass function for low-density supercluster-like groups 
identified with the linking density threshold $1+\delta>8$ shown in 
\S3 is indeed in accordance with the expectation of the excursion set theory.

Here we perform quantitative studies to investigate if the excursion set theory 
can be applicable to low-density supercluster-like groups. Particularly, we analyze if the filaments
defined in the excursion set theory of \citet{she06} have good correspondences to the low-density 
groups found in our simulations.     

For each group catalog found in our site percolation analyses with the linking density threshold
from $1+\delta=80$ for halos to $1+\delta=8$ below which the universality of the mass function
breaks down, its mass function is calculated and compared with the functional form of Eq. (3) derived from the excursion set theory.
It should be emphasized that the parameters in Eq. (3), $a$, $\alpha$ and $\beta$, are in
principle, not free parameters, but determined by the barrier of Eq. (8) given by dynamical collapse models. 
Thus twofolds of test should be included in the comparison of the theory against simulations,
namely the functional form of Eq. (3) itself, and the values of the parameters therein.
Here we first apply Eq. (3), regarding the parameters as free parameters, 
to fit the mass function from simulations. Then the barrier with the fitted values of the parameters
is compared with that expected from the ellipsoidal collapse model. In Figure 9, we show
the fitted barrier for groups with the linking density $1+\delta=16$, labeled as 'Fitting of original
MF". For comparison, we also show the Sheth-Tormen barrier for virialized halos. As expected,
the barrier for low density groups is lower and its slope with respect to $\sigma_0(M)$ is 
shallower than those of halos. However, we find that down to $1+\delta=8$, no group catalogs
have mass functions with fitted barriers that resemble the barrier of filaments given by Eq. (10) derived from 
the ellipsoidal collapse model in \citet{she06}. The theoretical barrier is nearly flat in 
$\sigma_0(M)$, while the fitted barriers all have significant slopes. The fitted positive slope seen in 
Figure 9 reflects that for the simulation results, the suppression of the mass function 
at low mass end relative to that at high mass end is stronger than that predicted by the excursion 
set theory with the barrier from the ellipsoidal collapse model. The discrepancy is clearly 
shown in Figure 10, where the green solid lines are the mass functions with $1+\delta=16$ for
all the simulations listed in Table 2, and the green dashed line is the mass function predicted
by the excursion set theory with a flat barrier. The disagreement between the simulation results and
the theoretical predictions can due either to the dynamical collapse model that gives rise to the 
barrier, or to the excursion approach itself. 

In the recent study of \citet{rob09}, they test the applicability of the excursion set theory for 
virialized halos against simulations. They obtain the collapse barrier directly from simulations
by tracing the collapse of virialized regions. They conclude that while the 
barrier is consistent with that from the ellipsoidal collapse model, 
the mass function from the excursion set theory with the obtained barrier is not in a good agreement with that  
from simulations. This indicates the existence of some intrinsic shortcomings in the excursion approach
itself. In \citet{ma10} and \citet{mag10a} they re-emphasize the importance of the non-Markovian corrections to the excursion set theory in predicting the halo mass function and halo bias for filters other than the sharp-k filter.  Within the spherical collapse model, an analytical formulation taking into account such corrections by introducing a $\kappa$ parameter is presented in Ma et al. (2010). Maggiore and Riotto (2010b) and Ma et al. (2010) also point out that the complicated halo formation process can be incorporated into a stochastic barrier to further improve the predictions of the excursion set theory. This results an additional parameter $a$ to change the barrier $\delta_c$ to $a^{1/2}\delta_c$ and $\kappa$ to $a\kappa$ in both the mass function and the bias for dark matter halos. 

Another problem known to the excursion set theory is that the mass function is derived based on  
the statistics of the trajectories of random points in Lagrangian space. On the other hand, 
the structure formation should happen mainly around peaks of the initial density fluctuation field. 
Consider a peak region that eventually forms a group of mass $M$. For a particle away from the central region of the peak, 
the gravitational interaction can drag
the particle into the group. However, the average linear density fluctuation 
at the position of that particle obtained by applying a spherical smoothing centered on itself 
over the scale corresponding to $M$ 
can be lower than the collapse barrier. Thus the particle is statistically 
assigned to lower mass groups in the excursion approach. This can lead to over predictions of 
low mass groups in comparison with those from simulations \citep[e.g.,][]{bon96, mon99, rob09}. 
It is expected that such an off-center problem can have more significant effects on low-density supercluster-like groups, 
such as filamentary groups, than those of high-density virialized halos. 
   
To overcome this, \citet{bon96} propose the peak-patch theory for halos.
In this approach, density peaks in the linear density fluctuation field smoothed over different
scales are found. Those peaks with heights above the collapse barrier are potential 
halo centers. The differential mass function at mass $M$ is then related to the 
derivatives of the number of peaks with respect to the smoothing scale at $M$. 
Such an approach can be regarded as the excursion approach only on peak particles.
The average excursion set theory on random particles in Lagrangian space is widely used because it is believed that
it should resemble statistically the peak excursion theory. 
However, to compare with the mass function from simulations, 
there is an important additional step in the peak excursion approach,  
namely, the peak exclusion, which is used to trim off overlapped peaks.
To certain extent, it is expected that the peak exclusion should largely remove the off-center problem
in the average excursion set theory discussed in the previous paragraph.
To see if this is the case, we perform the following analyses. Instead of trimming out peaks
in the excursion approach, we add in small-scale structures back to groups found in our simulations. 
Specifically, for each virialized halo identified with the linking density $1+\delta=80$,
we hierarchically increase $1+\delta$ from $80$ to $240$. At each level, if a new percolation group
occurs, and it is not the most massive subgroup of the parent group at the previous level, 
it is added into the original group catalog as an individual group. The same procedures are applied 
for supercluster-like groups identified with lower linking density thresholds. For example for groups with
the linking density threshold $1+\delta=16$, we hierarchically increase $1+\delta$ from $16$ to $80$
to add in subgroups back into the original group catalog.
We then analyze the mass functions of the modified group catalogs and compare them with those from the
excursion set theory of Eq. (3). In Figure 10, the mass functions for original halo catalogs and
the modified halo catalogs for all the simulations listed in Table 2 are shown in black and blue
solid lines, respectively. While the blue lines are systematically higher than the corresponding black lines
at low-mass ends, the two sets of mass functions are not significantly different. However, 
for low-density groups, the differences are large. The green and red lines in Figure 10
show the mass functions for original and modified group catalogs with the original linking density threshold $1+\delta=16$. 
The red lines are considerably higher than the green lines at low-mass ends. 
Fitting the modified mass functions to Eq. (3), we find that for the original linking density threshold $1+\delta=16$,
the fitted barrier is consistent with a flat-shaped barrier expected from the ellipsoidal collapse model of Eq. (10). 
The $a$ parameter for the amplitude of the barrier is found to be $a\approx 0.5$. This value of $a<1$  may be explained by the stochastic barrier model proposed by \citet{mag10b} and \citet{ma10}.
The fitted barrier is shown in Figure 9 labeled as "Fitting of modified MF", and the corresponding
mass function from Eq. (3) is shown by the red dashed line in Figure 10.
The good agreement between the red dashed line and red solid lines shows that taking into 
account the peak exclusion effects (in a reversed way here), the excursion set theory
with the barrier given by the ellipsoidal collapse model of \citet{she06} with adjusted $a\approx 0.5$ can describe well
the mass function of filaments. In this sense, it is appropriate for us to define groups identified 
with the linking density threshold $1+\delta=16$ as filaments.  

The analyses shown in this subsection demonstrate the importance of the peak exclusion effects
in the excursion set theory, especially when it is applied to model the mass function of 
low-density filamentary groups. Without considering such effects, the average excursion set theory
predicts significantly more low-mass filaments than those identified in simulations. 

\begin{figure}
\epsscale{1.0}
\plotone{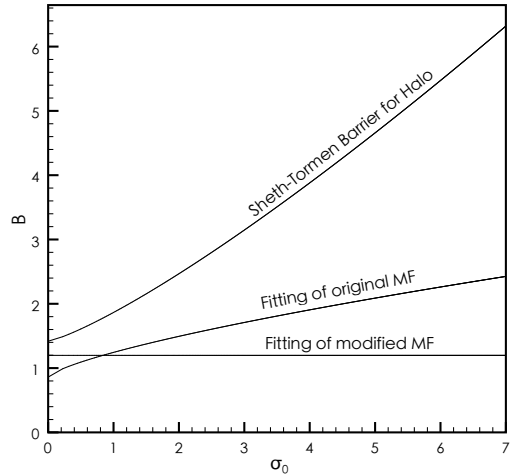}
\caption{Effective barriers for the excursion set theory. The Sheth-Tormen barrier for halo, 
and the fitted barriers for the original and modified groups with $1+\delta=16$ are shown.}
\end{figure}

\begin{figure}
\epsscale{1.0}
\plotone{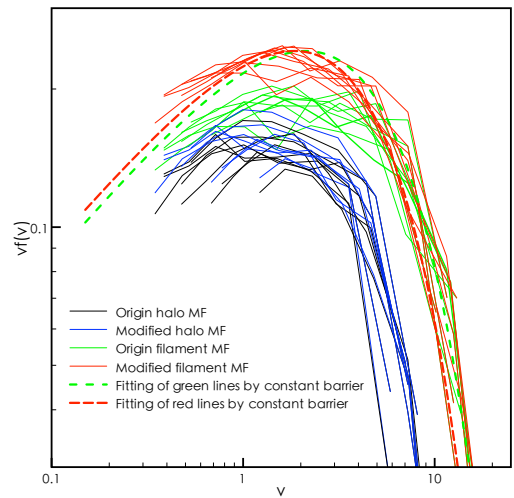}
\caption{Mass functions for the original and modified groups of halos $1+\delta=80$
and filaments ($1+\delta=16$) for simulations in Table 2 are shown. The heavy red and green dashed lines 
show the results from the excursion set theory [Eq. (3)] with flat barriers fitted to 
the red and green solid lines, respectively.}
\end{figure}

\begin{figure*}
\epsscale{1}
\plotone{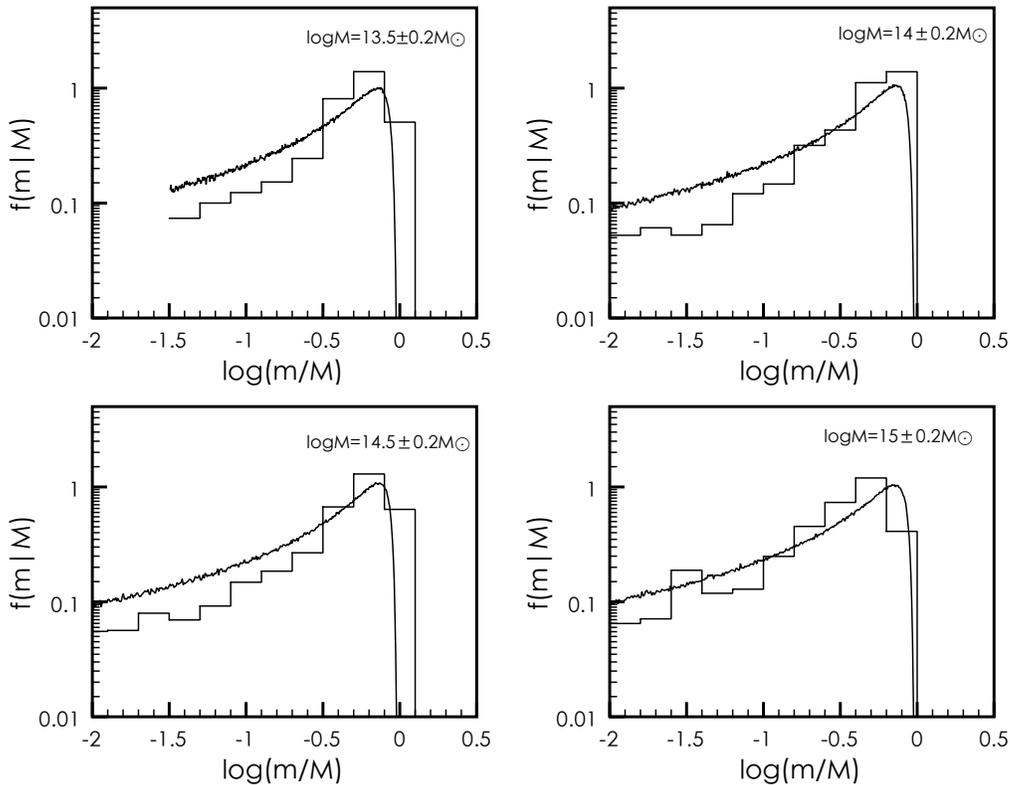}
\caption{Halo-filament conditional mass function $f(m|M)$. The histograms show results from simulation JS12,
and the solid lines with wiggles are for the results from the two-barrier excursion set theory obtained 
by sharp-$k$ Monte-Carlo simulations.} 
\end{figure*}

\begin{figure*}
\epsscale{1}
\plotone{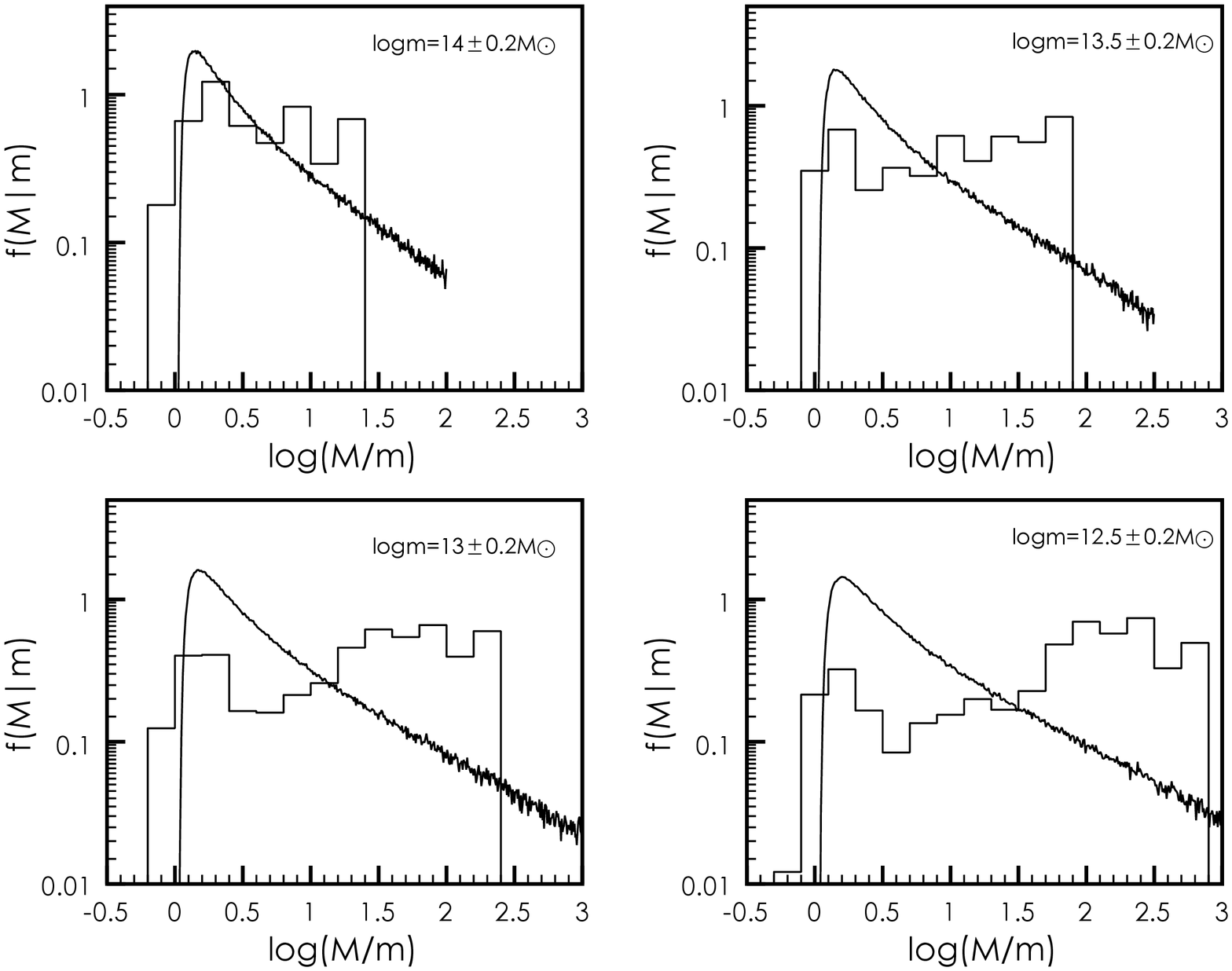}
\caption{Same as Fig. 11 but for filament-halo conditional mass function $f(M|m)$.} 
\end{figure*}

\subsection{Conditional Mass Function}

In the excursion set theory, the conditional mass function of objects can be analyzed by invoking two sets of barriers
that are appropriate for the two classes of structures in consideration.
The theory has been extensively applied to construct dark matter halo merging trees, where $f(m,z_1|M, z_2)$, the mass fraction of
the main halo of mass $M$ at redshift $z_2$ that is contained in progenitor halos of mass $m$ at redshift $z_1>z_2$,  
and $f(M,z_2|m, z_1)$, the probability that a halo of $m$ at redshift $z_1$ finds itself in halos of mass $M$ at $z_2<z_1$, 
are often investigated. In this case, the two barriers correspond to the barriers of halo formation at $z_1$ and $z_2$,
respectively \citep[e.g.,][]{lac93,kau93,som99,zha08}. In the framework proposed by \citet{she06} for the formation
of different types of objects, halos, filaments and sheets, the conditional mass functions 
can also be studied, which can potentially reveal the environmental dependence of structure formation.
Here we analyze the halo-filament and filament-halo conditional mass functions from simulations and compare them
with those predicted from the excursion set theory.
We denote $m$ as the mass of halos, and $M$ as the mass of filaments. Thus
$f(m|M)$ represents the halo-filament conditional mass function, i.e., the mass fraction of a filament of mass $M$
that is contained in halos of mass $m$.  
The filament-halo conditional mass function is written as $f(M|m)$, which 
gives the probability that a halo of mass $m$ locates at filaments of mass $M$. 

As shown in \S4.1, the mass function of filaments directly from simulations with the linking
density $1+\delta=16$ is not consistent with that from the nearly flat barrier derived by \citet{she06} for filaments. 
A moving barrier with a significant slope shown as "Fitting of the original MF" in Figure 9 is needed.
Then to test the validity of the excursion approach itself, 
in our calculations of the conditional mass functions from the excursion set theory, 
we do not use the barriers given by \citet{she06} for halos and filaments. Instead, 
the barriers with the parameters $a$, $\alpha$, and $\beta$ obtained by fitting Eq. (3) 
to the mass function of halos with $1+\delta=80$ and the mass function of filaments with $1+\delta=16$ from simulations
are adopted. For halos, the fitted barrier is very close to that of \citet{she01} shown in Figure 9. 
For filaments, the fitted barrier is the one labeled as "Fitting of the original MF" in Figure 9. 

Figure 11 shows $f(m|M)$, the halo-filament conditional mass function, for different values of $M$. 
The histograms are the results from simulations of JS12, and the solid lines are the sharp-k Monte-Carlo results
of the excursion set theory with the fitted barriers for halos and filaments. It is seen that the excursion set theory
can describe reasonably well the overall shape of $f(m|M)$. For relatively small $M$, 
it overestimates $f(m|M)$ at low $m/M$. This difference should not be due to numerical resolutions of the simulations
as we consider only halos with $m\ge 10^{12}\hbox{ M}_{\odot}$. Similar difference is also seen in 
\citet{col08} for halo-halo conditional mass function, where they compare the simulation results with those from the
extended Press-Schechter theory with the barriers from the spherical collapse model. 
Other studies indicate that taking into account the ellipsoidal collapse improves the agreement between the
the excursion set theory and simulations for the halo-halo conditional mass function \citep[e.g.,][]{gio07}.  
Note that in our analyses, the solid lines in Figure 11 are calculated by applying
the fitted barriers from the unconditional mass functions of halos and of filaments, respectively.
Thus the differences seen in Figure 11 should stem mainly from the excursion approach itself. 
The off-center problem discussed in \S4.1 can lead to an underestimation of merging probability
and thus an overestimation of the abundance of low-mass objects. Although this problem is not 
apparent for $f(m)$, the average halo mass function, it can have significant effects on the conditional mass function
$f(m|M)$ due to the relatively high-density environment. Furthermore, the underestimation of the
merging probability for filaments from the excursion set theory can have larger effects, leading to an 
overestimation of $f(m|M)$ at small $m/M$ especially for low-M filaments.

In Figure 12, we present the results for the filament-halo conditional mass function $f(M|m)$.
The simulation results shown as histograms appear to be rather extended. On the other hand, 
the results from the excursion set theory all have sharp peaks at $M/m\sim 1$. 
This should also be related to the off-center problem of the excursion approach. 
Given the halo mass $m$, filaments with $M\sim m$ most likely contain only single halos
and the halo centers overlap with their host filaments. 
Dynamically, those filaments are just
the extension of the halos therein. In such a case, the excursion approach can
describe well the conditional mass function $f(M|m)$ as expected. However, 
for filaments with $M>>m$, they normally have multiple halos in them, and
are likely the merging products of progenitor filaments each with 
a considered halo in it. Thus to understand
the behavior of $f(M|m)$, one needs to understand well the merging process of filaments. 
As we discussed earlier, the off-center problem in the excursion set approach 
is especially severe in explaining the formation of large filaments. 
Particularly, in analyzing the filament-halo conditional mass function
in which the existence of halos is pre-assumed, we are biased to emphasize regions with 
relatively high densities. These regions have high probabilities hosting 
large filaments. Therefore even with the fitted barrier from the unconditional mass function of filaments, 
the excursion set theory cannot predict well $f(M|m)$ at large $M/m$. 
Such a trend is also seen, albeit to a less extent, in the halo-halo conditional mass function
\citep[e.g.,][]{col08}. 

Because we use the effective barriers obtained by fitting to the unconditional mass functions of halos
and filaments from simulations, the differences between the excursion analyses and those from simulations
for both $f(m|M)$ and $f(M|m)$ are directly related to the differences in the joint probability distribution $f(m,M)$.
In simulations, merging process to form large filaments generates a valley in the $f(m,M)-(m,M)$ plot. 
This valley leads to the broad double-peak behavior for the filament-halo conditional mass function $f(M|m)$
that cuts $f(m,M)$ at a given $m$. For $f(m|M)$ which cuts $f(m,M)$ at a given $M$, 
the valley leads to a relatively rapid decrease of $f(m|M)$ at small $m/M$. 
On the other hand, the two-barrier excursion set theory cannot give rise to 
the valley, resulting the discrepancies seen both in Figure 11 and Figure 12. 
 
\section{Shape Statistics for filaments} 
In this section, we analyze the shape statistics of filament groups with $1+\delta=16$
to see if they indeed are filamentary like. 

\begin{figure*}
\epsscale{1.0}
\plotone{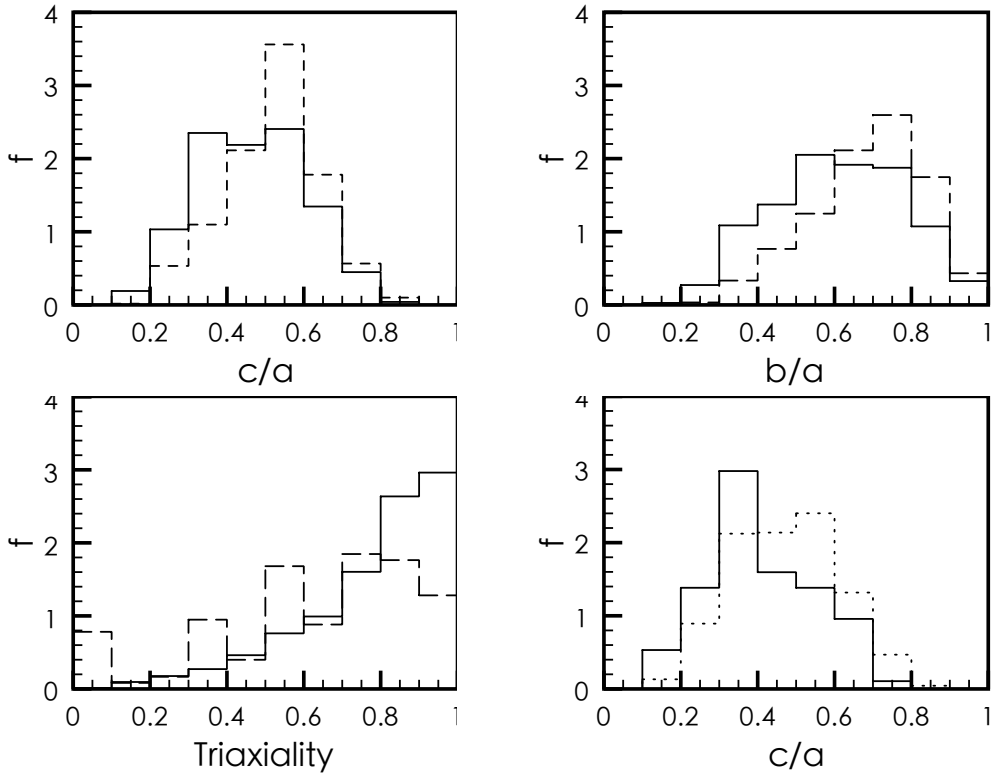}
\caption{Shape statistics for halos and filaments with mass larger than $10^{13}M_{\odot}$ for simulation JS12.
For the upper left, upper right and lower right panels, the dotted histograms are for halos, and 
solid histograms are for filaments. In the lower right panel, the solid and dotted histograms
are for filaments with $M>10^{14}M_{\odot}$ and $10^{13}M_{\odot}<M<10^{14}M_{\odot}$, respectively.}
\end{figure*}

We define the shape of a group through its grid-based inertia tensor, which is given by 
\begin{equation}
I_{i,j}=\sum_{n}x_{i,n} x_{j,n}
\end{equation} 
where the sum is over all grids occupied by a group and $\vec x$ is the grid central position with $i,j=1,2,3$. 
Note that we do not apply density weights to grid positions in calculating $I_{i,j}$. Thus our measurement
reflects the shape of the overall spatial extension of a group, and the contaminations from substructures
are minimal. By modeling approximately the spatial distribution of a group as a triaxial object,
we can link the axial ratios of the group to the eigenvalues $(\lambda_{a},\lambda_{b},\lambda_{c})$ of $I_{i,j}$ by 
$(a,b,c)=\sqrt{\lambda_{a},\lambda_{b},\lambda_{c}}$, where $a\ge b\ge c$ are the three axes.
Following \citet{fra91}, we define the triaxiality as 
\begin{equation}
T=\frac{a^2-b^2}{a^2-c^2}.
\end{equation}
An object is classified as an oblate object if its $T<1/3$, a traxial object if $1/3< T <2/3$, 
and a prolate object if $T>2/3$. 

Figure 13 presents the shape statistics. The upper two panels and the lower left panel are
for the axial ratios $c/a$, $b/a$ and the triaxiality $T$, respectively, for filament groups identified 
with the linking density threshold $1+\delta=16$ (solid histograms) and for halos with the linking density threshold 
$1+\delta=80$ (dashed histograms). It is observed that filaments tend to have smaller $c/a$ and $b/a$ than those of
halos. We have $c/a\sim 0.45$ and $b/a\sim 0.6$ for filaments, in comparison with
$c/a\sim 0.55$ and $b/a\sim 0.75$ for halos.
The shape differences between filaments and halos are best seen in the triaxiality $T$ statistics. 
For filaments, they are dominantly very prolate with $T>0.8$, consistent with the 
expected configuration for filaments. For halos, the distribution of $T$
is rather wide, and most of them are traxial in shape. The lower right panel shows the distribution of 
$c/a$ for filaments of two mass ranges, $M>10^{14}\hbox{ M}_{\odot}$ (solid) 
and $10^{13}\hbox{ M}_{\odot}<M<10^{14}\hbox{ M}_{\odot}$ (dotted), respectively.
The differences are apparent with $c/a\sim 0.35$ and $c/a\sim 0.5$ for high and low mass filaments,
respectively. From Figure 1, we can see that the massive filaments in the lower middle panel
locate at the major nodes of the global cosmic web (see for example, the upper right panel of Figure 1).
They usually contain multiple large halos, and their spatial extension reflects the spatial 
orientation of the global web structure. Dynamically, along the direction of the major axis
of these massive filaments, collapse has not happened yet. Thus to a large extent, 
these filaments should correspond closely to the two-axis collapsed filaments defined from linear density fluctuations. 
For relatively low mass filaments, most of them are extended structures of individual halos of similar mass. 
Therefore their shape distribution should be in accordance with that of halos. On the other hand,
because they include extended and dynamically unrelaxed regions, they tend to be somewhat more 
filamentary-like than halos therein.

\section{Summary and Discussion}

Applying a grid-based site percolation method to numerical simulations, we study groups identified with
different linking density thresholds $1+\delta$. Groups with $1+\delta=80$
correspond well to FOF dark matter halos. Lowering $1+\delta$ allows us to find supercluster-like
groups beyond virialized dark matter halos. As the linking density threshold approaches the average density
of the universe, the global cosmic web structure can be naturally found. 
In the studies presented in this paper, we focus on supercluster-like groups, which 
are expected to be dynamically bound, although not virialized yet. These groups
provide immediate environments to dark matter halos therein. Therefore understanding their properties
is an important step towards understanding the environmental effects on the formation and evolution of 
galaxies. 

Our analyses reveal that similar to dark matter halos, the mass functions of 
supercluster-like groups for different simulations listed in Table 2
also follow a universal behavior. This universality is consistent with the
consideration that these groups are gravitationally bound systems, and
form mainly through their own gravitational interactions. In other words, 
the universality found for supercluster-like groups and that for dark matter halos
should arise from the same origin. We further find that the Jenkins functional form 
can describe well the mass functions for not only halos, but also supercluster-like groups.
An extended Jenkins mass function applicable to both halos and supercluster groups
is then explicitly presented, in which the parameters $a$, $b$, and $c$ 
depend on the linking density threshold $1+\delta$. As expected, the universality of the mass functions
breaks down for groups with the linking density $1+\delta\le 8$ where the global web structures
occur. 

We also compare the mass functions from simulations with those from the excursion set theory
with effective barriers derived from the ellipsoidal collapse model. For halos with $1+\delta=80$, 
consistent with other studies, the two agree very well with the parameter $a$ adjusted to be 
$a=0.707$ for the moving barrier. For supercluster-like objects, the ellipsoidal collapse model
gives rise to a nearly flat barrier for filaments defined as two-axis collapse objects \citep[e.g.][]{she06}.
However, incorporating this barrier into the excursion set theory predicts a mass function
that cannot fit to any mass function of supercluster-like groups identified in simulations
with the linking density threshold $1+\delta<80$. The off-center problem in the excursion set theory 
leads to a significant over prediction for the mass function at low mass end. 
Taking into account this problem in the comparison, we find that the mass function of the groups identified with
$1+\delta=16$ is in good agreement with that from the excursion set theory for two-axis collapse filaments.
Defining these groups as filaments, we further study the halo-filament and filament-halo conditional
mass functions. Deviations from the predictions of the two-barrier excursion set theory are seen, 
which are especially significant for filament-halo conditional mass function. 

The studies carried out in this paper can have important cosmological applications. 
The universality of the mass functions found for supercluster-like groups raises 
a possibility for us to probe cosmologies with supercluster abundances. It can also be applied
to model statistically how the projection effects affect clusters' weak-lensing signals.  
In the very recent paper by \citet{mur10}, they identify filamentary galaxy groups from the 2dfGRS survey
using galaxy FOF method, and compare the properties of the groups with those of mock surveys constructed 
from numerical simulations. This study shows that it is becoming feasible observationally to 
analyze filamentary galaxy groups statistically, and they are in turn can be used as cosmological probes.  
Physically, we expect that these filamentary galaxy groups should be closely associated with the supercluster-like dark matter
groups in our studies. To quantitatively understand the relation between the two, detailed FOD modeling 
for galaxies in supercluster-like dark matter groups is necessary. We discuss the FOD for subhalos
in \S 3.2. For galaxies, however, the FOD can be much more complicated, and thorough 
investigations are highly desired.  
 
 It is further noted that analyses on real galaxy groups 
can only be done in redshift space. Redshift distortions from peculiar
velocities of galaxies can affect group identifications, and further
their mass functions and shape statistics considerably. For 
supercluster-like groups, their ambient member galaxies tend to 
be in the stage of coherent infall, and thus their distribution suffers oblate distortions in redshift space.
On the other hand, for their virialized inner regions, the distortion can  
generate finger-of-God structures in redshift space. 
The detailed impacts of redshift distortions on entire supercluster-like groups 
will be explored in our future studies.    
  
\acknowledgments
We are very grateful for Hou-Jun Mo for stimulating discussions, 
Yipeng Jing for kindly providing us his simulation data,
and Liang Gao and Jie Wang for their help on Virgo and GIF data. 
Virgo simulations and GIF simulations used in this paper were carried out by the 
Virgo Supercomputing Consortium using computers based at Computing Centre of 
the Max-Planck Society in Garching and at the Edinburgh Parallel Computing Centre. 
Their data are publicly available at www.mpa-garching.mpg.de/NumCos. 
Part of our analyses is done on SGI Altix 330 system at the Department of Astronomy, 
School of Physics, Peking University (PKU) and CCSE-I HP Cluster of PKU.  
This research is supported in part by the NSFC of China under grants 10373001, 10533010 and 10773001, and the 973
program 2007CB815401. 

{\it Facilities:} \facility{PKU}.

\end{document}